# Spatial imaging of the spin Hall effect and current-induced polarization in two-dimensional electron gases

V. Sih, R. C. Myers, Y. K. Kato, W. H. Lau, A. C. Gossard & D. D. Awschalom*

*Center for Spintronics and Quantum Computation, University of California, Santa Barbara, California 93106, USA*

**Spin-orbit coupling in semiconductors relates the spin of an electron to its momentum and provides a pathway for electrically initializing and manipulating electron spins for applications in spintronics[1] and spin-based quantum information processing[2]. This coupling can be regulated with quantum confinement in semiconductor heterostructures through band structure engineering. Here we investigate the spin Hall effect[3,4] and current-induced spin polarization[5,6] in a two-dimensional electron gas confined in (110) AlGaAs quantum wells using Kerr rotation microscopy. In contrast to previous measurements[7,8,9,10], the spin Hall profile exhibits complex structure, and the current-induced spin polarization is out-of-plane. The experiments map the strong dependence of the current-induced spin polarization to the crystal axis along which the electric field is applied, reflecting the anisotropy of the spin-orbit interaction. These results reveal opportunities for tuning a spin source using quantum confinement and device engineering in non-magnetic materials.**

Recent measurements in bulk epilayers of *n*-GaAs and *n*-InGaAs[7] and in a two-dimensional hole gas[8] provide experimental evidence for the spin Hall effect[3,4,11,12], but it remains unclear whether the dominant mechanism is extrinsic or intrinsic. The extrinsic mechanism[3,4] is mediated by spin-dependent scattering, where spin-orbit coupling mixes the spin and momentum eigenstates. Alternatively, an intrinsic spin Hall mechanism has recently been proposed[11,12] that is an effect of the momentum-dependent internal magnetic field $B_{int}$. This internal field arises from spin-orbit coupling, which introduces a spin splitting for electrons with non-zero wave vector **k** in semiconductors lacking an inversion center. For example, bulk inversion asymmetry exists due to the zincblende crystal structure of GaAs and introduces the Dresselhaus spin splitting[13], whereas structural inversion asymmetry is present in heterostructures that are not symmetric along the growth direction and leads to an in-plane spin splitting known as the Bychkov-Rashba effect[14]. The observation of the spin Hall effect in unstrained *n*-GaAs, in which the **k**-linear effective field is small[15], suggests that the extrinsic effect is dominant in that system[7]. However, a recent theoretical work argues that the cubic Dresselhaus term in GaAs could produce a non-negligible intrinsic spin Hall effect[16].

Measurements in (110) quantum wells (QW) may help distinguish between the two proposed mechanisms by allowing one to isolate the contributions of the Dresselhaus and Bychkov-Rashba fields. In two-dimensional systems, quantum confinement modifies the Dresselhaus spin splitting[17]. For (110) QW, the Dresselhaus field is oriented along the growth direction, whereas this field is in-plane in conventional (001) heterostructures. Since the Dresselhaus and Bychkov-Rashba fields are mutually perpendicular, one can tune the in-plane $B_{int}$ with the Bychkov-Rashba effect and the out-of-plane $B_{int}$ with the Dresselhaus field using engineered (110) heterostructures. In addition, two-dimensional systems provide a flexible architecture where carrier density, mobility, and structural inversion asymmetry can be controlled using electric fields[18].

Modulation-doped digitally-grown single QW are grown by molecular beam epitaxy on (110) semi-insulating GaAs substrates. The QW structure behaves like a single 75 Å $Al_{0.1}Ga_{0.9}As$ QW with $Al_{0.4}Ga_{0.6}As$ barriers at $T = 30$ K. For the optical measurements, a mesa is defined using a chemical etch (Fig. 1(a)), and contacts are made using annealed AuGe/Ni.

The spin polarization in the two-dimensional electron gas (2DEG) is spatially resolved using low temperature scanning Kerr rotation (KR) microscopy[19] in the Voigt geometry. A linearly polarized beam is tuned to the absorption edge of the QW ($\lambda = 719$ nm) and directed normal to the sample through an objective lens, providing ~1.1 μm lateral spatial resolution. The rotation of the polarization axis of the reflected beam provides a measure of the electron spin polarization along the beam direction. A square wave voltage with maximum amplitude $\pm V_p$ and frequency 511 Hz is applied to the device for lock-in detection. Measurements are performed in devices with electric fields applied along four different crystal directions in order to create a directional map of the internal fields. All of the data presented are measured at $T = 30$ K, and we take $x = 0$ μm to be the center of the channel.

In Fig. 1(b), we present KR data as a function of the applied in-plane magnetic field $B_{ext}$ for positions near the two opposite edges of a channel aligned along the [001] direction. This data corresponds to a measurement of the Hanle effect using KR[20] and indicates the presence of an out-of-plane spin polarization when the data can be fit to a Lorentzian $A_0/[(\omega_L \tau_s)^2 + 1]$, where $A_0$ is peak KR, $\omega_L = g \mu_B B_{ext}/\hbar$ is the Larmor precession frequency, $\tau_s$ is the electron spin coherence time, $g$ is the electron g factor, $\mu_B$ is the Bohr magneton, and $\hbar$ is the Planck constant. $A_0$ is of opposite sign for the two edges of the sample, which is a signature of the spin Hall effect.

In Fig. 1(c), a one-dimensional spatial profile of the spin accumulation near the edges is mapped out by repeating $B_{ext}$ scans as a function of position. There are two spin Hall peaks at each edge, one around $x = \pm 58.6$ μm and one of smaller amplitude around $x = \pm 56.4$ μm. This structure was not observed in measurements on bulk epilayers[7], and could be due to an additional contribution from spin-polarized carriers undergoing spin precession about the in-plane Bychkov-Rashba field as they diffuse towards the center of the channel. However, the asymmetry in $|A_0|$ for the right and left edges and a spatial dependence of $\tau_s$ was also observed in previous measurements[7]. The reflectivity R shows the position of the edges of the channel, at $x = \pm 59.4$ μm.

In the [001]-oriented device, electrically-induced spin polarization is observed only at the edges of the channel. In contrast, devices fabricated along the [1$\bar{1}$0], [1$\bar{1}$1], and [$\bar{1}$12] directions also exhibit spin polarization at the center of the channel. Fig. 2(b) shows data taken at $x = 0$ μm for **E** along [1$\bar{1}$0], [1$\bar{1}$1] and [$\bar{1}$12]. Since the polarization is along the growth direction and depends on the direction of **E** relative to the crystal axes, we attribute this effect to the Dresselhaus field. The application of an electric field results in a non-zero average drift velocity of the electrons, which produces a non-zero effective magnetic field that orients spins[5,6]. Although the opposite sign of $A_0$ for **E** || [1$\bar{1}$0] and **E** || [1$\bar{1}$1] may seem surprising since these directions are only separated by 35.3° in the (110) plane (Fig. 2(a)), it is consistent with the calculated $B_{int}$ due to the cubic Dresselhaus field in a (110) QW. This theory also predicts that $B_{int}$ should be zero for **E** || [001][21], as observed.

Fig. 2(c) shows a spatial profile of the spin polarization near the edges for a device aligned along [1$\bar{1}$0]. $A_0$ is negative across the entire channel, and $|A_0|$ increases with increasing voltage. From -52 μm < $x$ < +52 μm, $|A_0|$ is nearly constant across the channel. However, $|A_0|$ becomes smaller near the left edge of the channel, and a negative peak in $A_0$ is evident near the right edge, which is due to the spin Hall effect. The data for $V_p = 3$ V suggests that there may be two spin Hall peaks, at $x = 55.5$ μm and $x = 57.5$ μm, which is similar to the two peaks with ~2 μm spacing observed in the [001] device. We also observe that $A_0$ increases more dramatically with voltage for the spin Hall peak near the right edge than for the current-induced spin polarization across the rest of the channel.

We continue examining the direction dependence of the current-induced spin polarization with spatial scans of a channel aligned along [1$\bar{1}$1]. Figure 3 shows the spatial profile of the spin polarization near the edges of the channel. $A_0$ is positive across the entire channel, and $|A_0|$ is nearly constant from -26 µm < $x$ < +26 µm. However, there is a small positive peak around $x$ = -31 µm, and $|A_0|$ diminishes near the right edge of the channel.

We also perform spatially-resolved measurements of a device aligned along $[\bar{1}12]$ (Supplementary Figure 1). Again, we observe a uniform spin polarization in the center of the channel and spin accumulation due to the spin Hall effect at the edges of the channel. From our measurements on all four devices, we conclude that the spin Hall effect exhibits the same polarity for electric fields applied along all four crystal directions.

In Fig. 4, we present voltage dependences of $A_0$ and $\tau_s$ for the spin Hall peaks in the [001] device and the current-induced spin polarization in the $[1\bar{1}0]$, $[1\bar{1}1]$ and $[\bar{1}12]$ devices. In Fig. 4(a), we plot $A_0$ for the spin Hall peaks near the edges of the [001] channel and observe that $|A_0|$ increases with increasing $V_p$. The non-linearity of the increase in $|A_0|$ could be due to changes in the spin Hall profile or in the electrical response of the device. In contrast, we observe in Fig. 4(b) that $\tau_s$ = 545 ± 176 ps and does not have a clear voltage dependence over this range.

In order to explore the direction dependence of the current-induced spin polarization, we measure $A_0$ at $x$ = 0 µm for devices aligned along $[1\bar{1}0]$, $[1\bar{1}1]$, and $[\bar{1}12]$ as a function of $V_p$, which we plot in Fig. 4(c). We observe that the amplitude of the current-induced spin polarization increases with increasing $V_p$, as expected. In addition, $\tau_s$ = 1344 ± 404 ps and does not exhibit a clear dependence with voltage (Fig. 4(d)). The direction dependence of $A_0$ reflects the strong **k**-dependence of the Dresselhaus field.

In order to determine the mechanism of the spin Hall effect, we quantify the Rashba coefficient α by measuring the in-plane $B_{int}$ for our sample. The Bychkov-Rashba field has magnitude $|B_{int}| = α |k| / g μ_B$ and is oriented perpendicular to **k**. $B_{int}$ can be observed as a shift in a Hanle[22] or field-dependent KR curve[15] when we apply a DC voltage $V_{DC}$ along the [001] direction. Spins are injected optically into the QW and measured as a function of $B_{ext}$ after a time delay of 6 ns. Figure 5(a) shows KR as a function of $B_{ext}$ for $V_{DC}$ = -2 V and $V_{DC}$ = +2 V. Lorentzian fits determine the center of the peak, which is -$B_{int}$. In Fig. 5(b), $B_{int}$ as a function of $V_{DC}$ can be fit to a line with slope 1.77 mT/V, and we determine α = 1.8 x $10^{-12}$ eV m. This small value for α is reasonable since this QW was designed to be symmetric, as α is a measure of the structural inversion asymmetry. This is also consistent with the observation that the current-induced spin polarization does not change significantly for the **E** || $B_{ext}$ geometry, where one would also measure spins that are oriented in-plane[9]. In addition, this value for $B_{int}$ yields a spatial spin precession period[23] of 3.5 μm, which is similar to the ~2 μm distance observed between the spin Hall peaks in the [001] and [1$\bar{1}$0] devices and suggests that the spacing between the spin Hall peaks could be due to spin precession. This relation could be confirmed by tuning α with a gate-voltage[18]. Calculations of the intrinsic spin Hall effect for Rashba spin-orbit coupling show that the spin Hall conductivity should be non-zero when the Rashba splitting is larger than the disorder broadening[12]. The ratio $\frac{\Delta_0 \tau_p}{\hbar} \sim 10^{-6}$, where $\tau_p$ is the mean scattering time, and relates the strength of the spin-orbit coupling with impurity scattering[24]. In addition, the Dresselhaus terms are oriented out-of-plane in our sample and should not contribute to the spin Hall conductivity. Therefore, our data suggests that the spin Hall effect that we observe is dominated by the extrinsic spin Hall mechanism.

Spin-orbit engineering in two-dimensional systems allows for the manipulation of the magnitude and direction of the internal fields for sourcing spin polarization in non-magnetic semiconductors. Moreover, these interactions can be used to operate on electron spins by changing the direction of current, thereby enabling new degrees of control for quantum confined spintronic devices.

**Methods**

Sample growth and device preparation

Conditions for the (110) growth are similar to those described in Ref. [25]; the substrate temperature is 490°C, the As$_4$ beam equivalent pressure is 1.6 x 10$^{-5}$ torr, and the growth rate of GaAs is ~0.5 ML/sec. The samples consist of four 14 Å GaAs layers with Al$_{0.4}$Ga$_{0.6}$As barriers separated by 6 Å Al$_{0.4}$Ga$_{0.6}$As spacers. The barriers are delta-doped with Si at 200 Å from the QW structure on both the surface and the substrate side, with doping densities of 1.4 x 10$^{12}$ cm$^{-2}$ and 5.6 x 10$^{11}$ cm$^{-2}$, respectively. In addition, Si-doping at 1 x 10$^{18}$ cm$^{-3}$ is present within the QW region. Conventional Hall measurements at a temperature $T = 5$ K determine the sheet density $n_s = 1.9$ x 10$^{12}$ cm$^{-2}$ and mobility $\mu = 940$ cm$^2$/V s. Devices are aligned to the natural cleaves along [001] and [1$\bar{1}$1] such that an electric field **E** can be applied along the in-plane directions [001], [1$\bar{1}$0], [1$\bar{1}$1], and [$\bar{1}$12]. Using time-resolved KR[26], we determine $|g| = 0.33$ for this sample, and $\tau_s = 766$ ps at B$_{ext} = 0.2$ T. The longitudinal spin coherence time is 3250 ps at B$_{ext} = 0$ T. The relatively long spin coherence times observed in (110) QW[27] compared to (001) 2DEGs[28] is due to the suppression of the D'yakonov-Perel' spin relaxation mechanism[29]. The data presented in this paper are from devices processed from one sample, but measurements performed on devices fabricated from a second sample verify the reproducibility of our results.

Measurement of Bychkov-Rashba field

The shift in field-dependent KR is used to measure the in-plane B$_{int}$ as a function of applied voltage in order to determine α. Since the contact resistance is large compared to the resistance of the channel, we consider the voltage drop across the channel

$$V_c = \frac{R_c}{R_T} V_{DC},$$ where $R_c = 980 \ \Omega$ is the resistance of the channel and $R_T = 10.3 \ k\Omega$ is the total resistance of the device. Since $\langle \mathbf{k} \rangle = \frac{\mu V_c m_e^*}{\hbar l}$, where the in-plane effective mass $m_e^* = 0.074 \ m_e$ from a 14-band $\mathbf{K} \cdot \mathbf{p}$ calculation, and the spin splitting energy $\Delta_0 = g\mu_B B_{int}$, we determine $\alpha = 1.8 \ \mathrm{x} \ 10^{-12}$ eV m.

**Figure 1.** Spin Hall effect in a two-dimensional electron gas (a) Device schematic and measurement geometry. The light blue region indicates the mesa, and the yellow regions are the contacts. (b) Kerr rotation (hollow symbols) and fits (lines) as a function of applied in-plane magnetic field $B_{ext}$ for $x$ = -58.4 µm (top, in red) and $x$ = +58.4 µm (bottom, in blue). The channel has width $w$ = 120 µm, length $l$ = 310 µm, and mesa height $h$ = 0.1 µm. A linear background is subtracted for clarity. (c) $B_{ext}$ scans as a function of position near the edges of the channel of a device fabricated along $[001]$ for $V_p$ = 2 V. Amplitude $A_0$, spin coherence time $\tau_s$, and reflectivity $R$ are plotted for $V_p$ = 1.5 V (blue squares) and 2 V (red circles).

**Figure 2.** Current-induced spin polarization in a two-dimensional electron gas (a) Relative orientations of crystal directions in the (110) plane. (b) Kerr rotation (hollow symbols) and fits (lines) as a function of $B_{ext}$ for **E** || $[1\bar{1}0]$ (black), **E** || $[1\bar{1}1]$ (red), and **E** || $[\bar{1}12]$ (green) at the center of the channel. (c) $B_{ext}$ scans as a function of position near the edges of the channel of a device fabricated along $[1\bar{1}0]$ with $w$ = 118 µm and $l$ = 310 µm for $V_p$ = 2 V. Amplitude $A_0$, spin coherence time $\tau_s$, and reflectivity $R$ are plotted for $V_p$ = 1.5 V (blue squares), 2 V (red filled circles) and 3 V (black open circles).

**Figure 3.** Spin polarization near the edges of a channel oriented along $[1\bar{1}1]$. $B_{ext}$ scans as a function of position near the edges of the channel of a device fabricated along $[1\bar{1}1]$ with $w$ = 68 µm and $l$ = 306 µm for $V_p$ = 1.5 V. Amplitude $A_0$ and reflectivity $R$ are also plotted.

**Figure 4.** Voltage dependence of the electrically-induced spin polarization. (a) Amplitude $A_0$ and (b) spin coherence time $\tau_s$ of the spin Hall polarization as a function of voltage for $x = -58$ μm (red) and $x = +58$ μm (blue) for a device fabricated along $[001]$. (c) $A_0$ and (d) $\tau_s$ of the current-induced spin polarization as a function of voltage $V_p$ for electric fields applied along $[1\bar{1}0]$ (black), $[1\bar{1}1]$ (red), and $[\bar{1}12]$ (green) measured at the center of the channel ($x = 0$ μm).

**Figure 5.** Measurement of the Bychkov-Rashba spin splitting. (a) Kerr rotation as a function of $B_{ext}$ for $V_{DC} = -2$V (blue) and $V_{DC} = +2$ V (red). The data was taken with a laser spot size of 30 μm. Lines are Lorentzian fits. (b) In-plane effective magnetic field $B_{int}$ as a function of $V_{DC}$.

**Supplementary Figure 1.** Spin polarization across a channel oriented along $[\bar{1}12]$. $B_{ext}$ scans as a function of position near the edges of the channel of a device fabricated along $[\bar{1}12]$ with $w = 60$ μm and $l = 306$ μm for $V_p = 2$ V. Amplitude $A_0$ and reflectivity $R$ are also plotted.

**Acknowledgments.** We acknowledge support from ARO, DARPA, NSF, and ONR.

Correspondence and requests for materials should be addressed to D. D. A. (e-mail: awsch@physics.ucsb.edu).

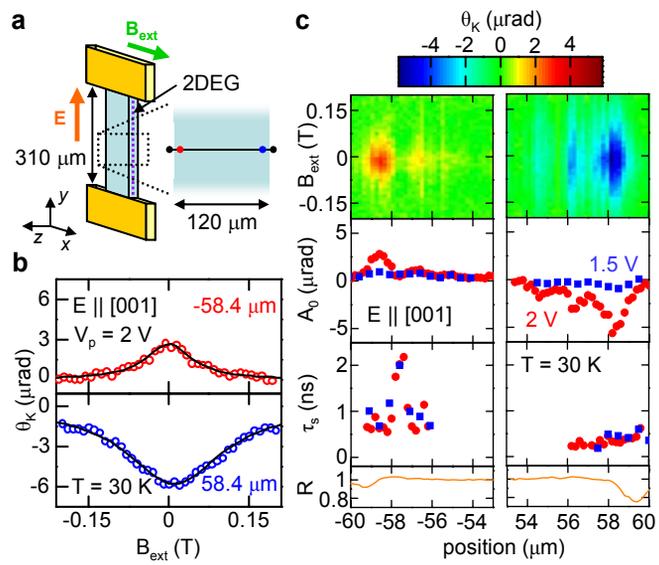

Figure 1
Sih *et al.*

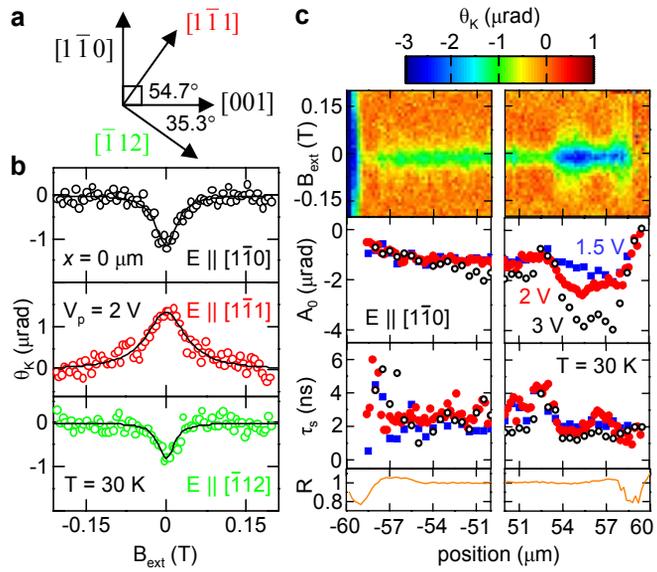

Figure 2
Sih *et al.*

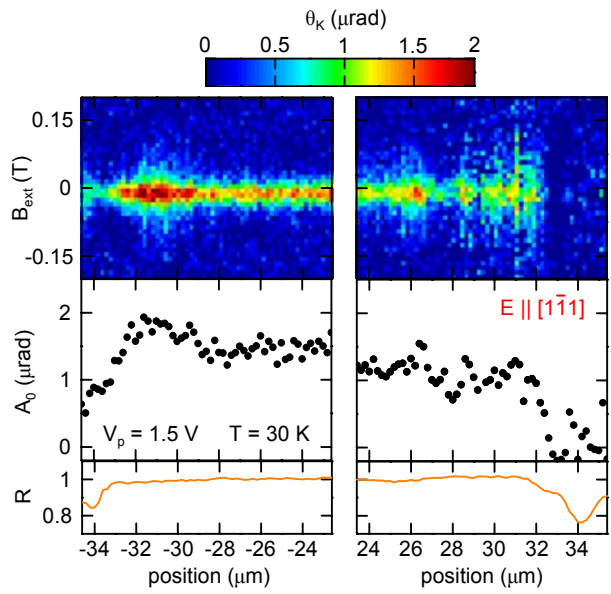

Figure 3
Sih *et al.*

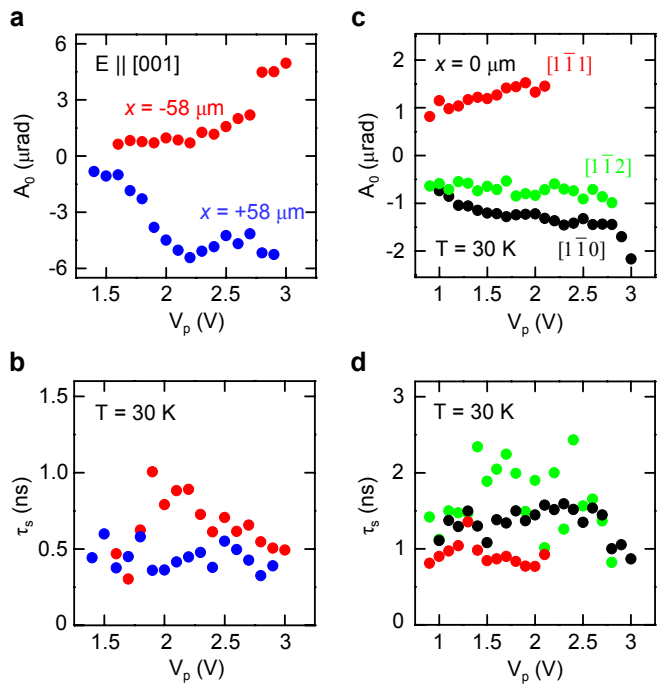

Figure 4
Sih *et al.*

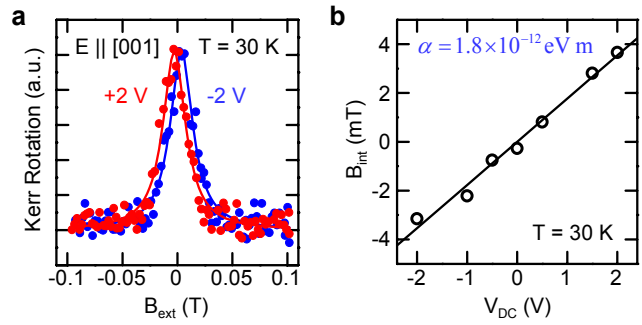

Figure 5
Sih *et al.*

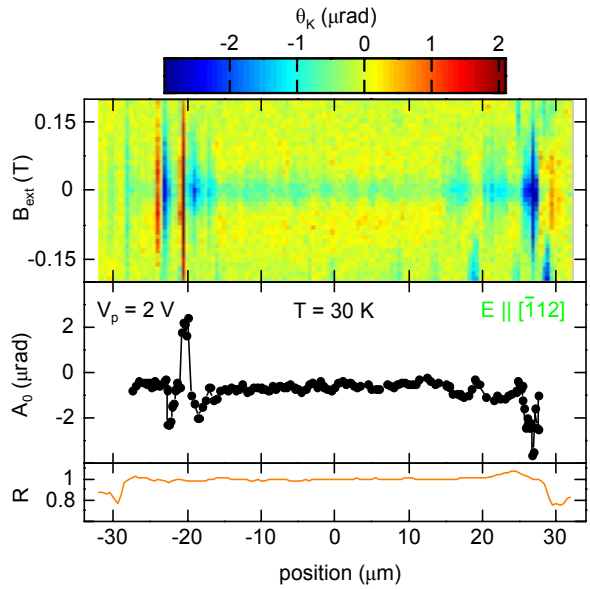

**Supplementary Figure 1.** Spin polarization across a channel oriented along $[\bar{1}12]$. $B_{ext}$ scans as a function of position near the edges of the channel of a device fabricated along $[\bar{1}12]$ with $w = 60$ μm and $l = 306$ μm for $V_p = 2$ V. Amplitude $A_0$ and reflectivity $R$ are also plotted.